\documentclass{iaus}
\usepackage{graphicx}

\title[Beryllium in the halo and the thick disk] 
{Beryllium abundances and the formation of the halo and the thick disk}

\author[Smiljanic et al.]   
{Rodolfo Smiljanic$^{1,2}$
  L. Pasquini$^2$,
  P. Bonifacio$^{3,4,5}$,
  D. Galli$^6$, \\
  B. Barbuy$^1$,
  R. Gratton$^7$,
 \and S. Randich$^6$
 }

\affiliation{$^1$IAG, University of S\~ao Paulo, S\~ao Paulo, Brazil,  
$^2$ESO, Garching bei M\"unchen, Germany, 
 \\ email: {\tt rsmiljan@eso.org} \\[\affilskip]
$^3$GEPI Observatoire de Paris - Meudon, France, 
$^4$INAF, Osservatorio di Trieste, Trieste, Italy, 
$^5$CIFIST Marie Curie Excellence Team, 
$^6$INAF- Osservatorio di Arcetri, Firenze, Italy
$^7$INAF-Osservatorio di Padova, Padova, Italy, 
}

\pubyear{2010}
\volume{268}  
\jname{Light elements in the Universe}
\editors{C. Charbonnel, M. Tosi, F. Primas \& C. Chiappini, eds.}
\begin{document}

\maketitle

\begin{abstract}
The single stable isotope of beryllium is a pure product of cosmic-ray spallation in the ISM. Assuming 
that the cosmic-rays are globally transported across the Galaxy, the beryllium production should 
be a widespread process and its abundance should be roughly homogeneous in the early-Galaxy at a
given time. Thus, it could be useful as a tracer of time. In an investigation of the use of Be as a 
cosmochronometer and of its evolution in the Galaxy, we found evidence that in 
a log(Be/H) vs.\, [$\alpha$/Fe] diagram the halo stars separate into two components. 
One is consistent with predictions of evolutionary models while the other is chemically 
indistinguishable from the thick-disk stars. This is interpreted as a difference in the star 
formation history of the two components and suggests that the local halo is not a single uniform 
population where a clear age-metallicity relation can be defined. We also found evidence that
the star formation rate was lower in the outer regions of the thick disk, pointing towards an 
inside-out formation. 
\keywords{stars: abundances -- stars: late-type -- Galaxy: halo -- Galaxy: thick disk}
\end{abstract}

\firstsection 
\section{Introduction}

The single stable isotope of beryllium, $^{9}$Be, is a pure product of cosmic-ray spallation 
of heavy nuclei (mostly CNO) in the interstellar medium \cite[(Reeves et al. 1970)]{Ree70}. 
In this sense, the production of Be can occur in two ways. In the so-called direct process  
the cosmic rays are composed of protons and $\alpha$-particles and these collide with CNO 
nuclei of the medium. In the so-called inverse process the cosmic rays are composed of 
accelerated CNO nuclei that collide with protons and $\alpha$-particles of the medium.

Observational works on Be abundances in metal-poor stars \cite[(Rebolo et al. 1988; Gilmore et al. 1992; 
Molaro et al. 1997; Boesgaard et al. 1999; Smiljanic et al. 2009)] {Reb88,Gil92,Mol97,Boe99,Sm09} find that log(Be/H) and
[Fe/H] (or [O/H]) show a linear relation with slope close to one. Such
slope argues that Be behaves as a primary element in the early Galaxy and its production
mechanism is independent of the ISM metallicity. This means that the
dominant production mechanism of Be is the inverse process \cite[(Duncan et al. 1992)]{Dun92}.

If one assumes that the cosmic-rays are globally transported across the Galaxy, than it 
follows that the Be production should be a widespread process. Beryllium can be produced anywhere 
in the Galaxy. One may thus expect that the Be
abundances are rather homogeneous at a given time in the early
Galaxy. It should have a smaller scatter than the products of stellar
nucleosynthesis \cite[(Suzuki et al. 1999; Suzuki \& Yoshii 2001)]{Suzuki99,SY01} and could thus 
be employed as a cosmochronometer for the early stages of the
Galaxy \cite[(Beers et al. 2000; Suzuki \& Yoshii 2001)]{Beers00,SY01}.

This was tested by \cite[Pasquini et al. (2004, 2007)]{Pas04,07} who calculated Be abundances in turn-off stars 
of two globular clusters, NGC 6397 and NGC 6752. These Be abundances were used to derive ages by means of a 
comparison with a model of the evolution of Be with time (\cite[Valle et al. 2002]{Valle02}). These ages agree well 
with those derived from theoretical isochrones, supporting the use of Be as a cosmochronometer. 

\cite[Pasquini et al. (2005)]{Pas05} then used a sample of 20 halo and thick-disk stars, 
previously analyzed by \cite[Boesgaard et al. (1999)]{Boe99}, to investigate the evolution of 
star formation rate in these two Galactic components. The idea is to use a diagram of [O/Fe] vs. log(Be/H) 
where the abscissa can be considered as increasing time and the ordinate as the star formation rate. 
A possible separation between stars of the two components was found and interpreted as a difference in the time 
scales of star formation.

\cite[Smiljanic et al. (2009)]{Sm09} analyzed the largest sample of halo and 
thick-disk stars to date, extending the investigation of Be as cosmochronometer and its role 
as a discriminator of the different stellar populations in the Galaxy. These results are discussed 
in more detail in the following sections.

More details about Be can also be found in the references cited above and in many contributions in 
this volume, e.g., A. Boesgaard; D. Lambert; F. Primas; and H. Reeves.

\section{The relation of Be with [Fe/H] and [$\alpha$/H]}

\begin{figure}[t]
\begin{center}
 \includegraphics[width=5.2in]{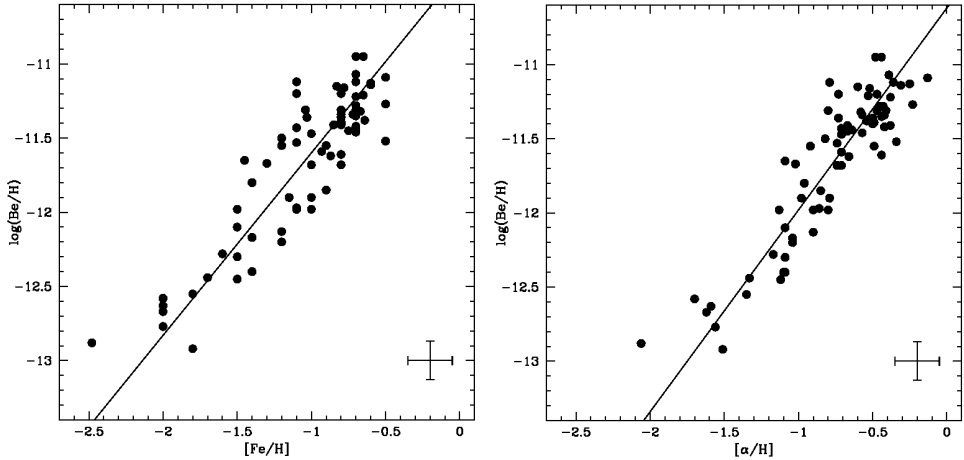} 
 \caption{Abundances of Be as a function of [Fe/H] (left panel) and of [$\alpha$/Fe]  (right panel). 
 This figure is adapted from \cite[Smiljanic et al. (2009)]{Sm09}.}
   \label{fig1}
\end{center}
\end{figure}

The beryllium abundances calculated by \cite[Smiljanic et al. (2009)]{Sm09} are shown in Fig. 1 
as a function of both [Fe/H] and [$\alpha$/H]. Linear relations with slopes close to one, as found 
in previous works, are seen in these plots (the slope is 1.23 for [Fe/H] and 1.37 for [$\alpha$/H]). 

The scatter of the abundances seen in these plots is larger than that found 
by previous works in the literature. Statistically, because of the size of the error bars, it is not 
possible to say whether the observed scatter is real. However, it was possible to find among the sample 
stars, stars that have similar atmospheric parameters, same metallicity, normal unaltered Li abundances, 
but different Be abundances. The direct comparison of the spectra of these stars is shown in Fig. 14 of 
Smiljanic et al. (2009). This comparison strongly argues that at least part of the observed scatter is real.

Two different explanations can be put forward to understand the scatter. One is that it is caused by 
local effects, such as the proximity to supernovae (or as suggested by \cite[Smiljanic et al. (2008)]{Sm08} 
to explain the case of the Be-rich star HD 106038 the proximity to a hypernova). The other is related to 
the stars belonging to different stellar populations. As we discuss below, this second explanation is 
the preferred one.

A figure similar to Fig. 1 with the stars divided according to its membership to the halo or the thick disk star 
does not show a obvious division. This may be caused in part by the narrow metallicity range of the 
thick-disk stars (metal-poor thick-disk stars are rare). However, if such a plot is made it becomes clear that 
the scatter of the abundances for the halo stars is larger than that for the thick-disk stars.

\section{Stellar Populations}

\begin{figure}[t]
\begin{center}
 \includegraphics[width=5.2in]{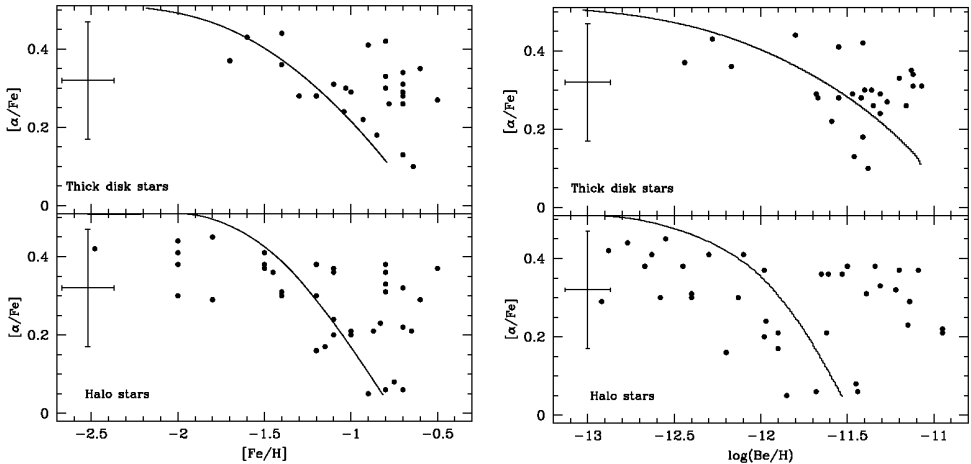} 
 \caption{Diagram of [$\alpha$/Fe] as a function of [Fe/H] (left panel) and of log(Be/H) (right panel). 
 The thick-disk stars are shown in the upper panels and the halo stars in the lower ones. The thick-disk 
 stars behave in the same way in both plots. The halo stars, however, clearly divide into two sequences 
 when [$\alpha$/Fe] is plotted against log(Be/H). The curves are the predictions of the models by 
 Valle et al. (2002). This figure is adapted from \cite[Smiljanic et al. (2009)]{Sm09}.}
   \label{fig2}
\end{center}
\end{figure}

\begin{figure}[t]
\begin{center}
 \includegraphics[width=2.6in]{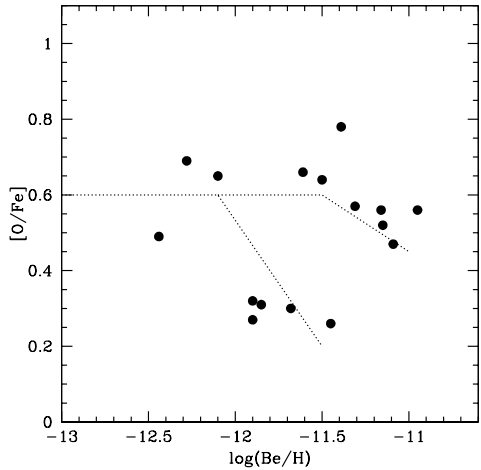} 
 \caption{The [O/Fe] ratio as a function of beryllium abundances for a subsample of the halo 
 stars analyzed in \cite[Smiljanic et al. (2009)]{Sm09}. These preliminary oxygen abundances 
 were calculated using the OI triplet at 777nm. The abundances are corrected for NLTE 
 (\cite[Fabbian et al. 2009]{Fabbian09}). The lines are only traced to guide the eye. }
   \label{fig3}
\end{center}
\end{figure}

We further investigate the different stellar populations in Fig. 2, using a diagram of [$\alpha$/Fe] vs. 
log(Be/H). In \cite[Smiljanic et al. (2009)]{Sm09} oxygen abundances were not available, 
so mean abundances of $\alpha$-elements were used instead. Again, the abscissa can be 
considered as increasing time and the ordinate as the star formation rate. Here we also 
present some new preliminary oxygen abundances determined from the infrared triplet at 777nm 
for a subsample of the halo stars (Fig. 3). The oxygen abundances were calculated using 
spectrum synthesis and the same codes and line lists used in \cite[Smiljanic et al. (2008, 2009)]{Sm08,Sm09}. 
The oxygen abundances are corrected for NLTE effects following \cite[Fabbian et al. (2009)]{Fabbian09}. 
To calculate the oxygen abundances a new reduction of the red-arm UVES spectra, where the OI 
lines are located, was necessary. The red UVES spectra is usually strongly affected by fringing. This can, 
however, be corrected with a new reduction using the latest UVES pipeline.

When using the Be abundances instead of Fe, it becomes clear that, with either $\alpha$ or oxygen abundances, the halo stars define 
two clear distinct sequences (Figs. 2 and 3). One sequence is chemically similar to the thick disk, the other 
agrees with the behavior expected for the halo stars in the models of \cite[Valle et al. (2002)]{Valle02}.  The 
thick disk stars behave the same no matter if using Be or Fe (Fig. 2).

Checking the kinematics of the stars, however, it is possible to notice that the group of low-$\alpha$ halo stars, 
that follows the expected behavior of the models, have similar kinematics. We define this group from the diagram 
of Fig. 2 as the stars with [$\alpha$/Fe] $\leq$ 0.25 and log(Be/H) $\leq$ $-$11.4. In Fig. 4 we show diagrams of 
both [$\alpha$/Fe] and log(Be/H) as a function of V, the component of the space
velocity of the star in the direction of the disk rotation, and of R$_{\rm min}$, the perigalactic distance of the
stellar orbit. The low-$\alpha$ group (open symbols) have mostly  V $\sim$ 0 and R$_{\rm min}$ $\leq$ 1 
kpc. They seem to be a group of non-rotating stars going very close to the Galactic center, a behavior that might be expected to
be shown by accreted stars that sink to the Galactic center by dynamical friction.  They also form a very tight and well-defined sub-sequence 
in a diagram of [Fe/H] vs. log (Be/H) (Fig. 5). This fact helps to explain why the halo stars show a larger 
scatter in this kind of diagram. 

The splitting into two components may be related to the accretion of external systems or to variations of star 
formation in different and initially independent regions of the early halo. The interpretation is still open, 
it is however clear that the halo is not a single uniform population with a single age-metallicity relation. 
A similar division was found by \cite[Nissen \& Schuster (1997, 2009)]{NS97,NS09} but using Fe as 
a tracer of time. The division seems clearer when Be is used as a time scale.

For the thick disk, it is possible to see in the lower-right panel a significant anticorrelation of [$\alpha$/Fe] with
perigalactic distance. This anticorrelation might be interpreted as evidence that the SFR was lower in the outer regions of 
the thick disk. A similar correlation however is not seen for Be. This lack of anticorrelation seems to indicate 
that Be is not affected by the local details of star formation, confirming that it can be used as a time scale. Although 
the range in Be abundances covered by the thick disk is very small, it is possible to see in Fig. 4 that the thick disk stars 
with smallest Be concentrate in the inner regions of the disk. As these are expected to be old stars, this result seems to 
point towards an inside-out formation of the thick disk.

\begin{figure}[t]
\begin{center}
 \includegraphics[width=5.2in]{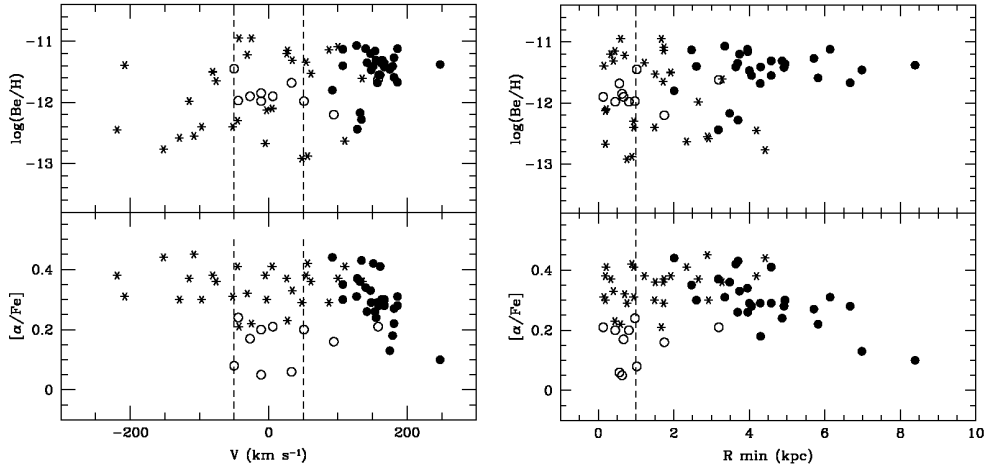} 
 \caption{The [$\alpha$/Fe] ratio and the Be abundances as a function of V, the component of the space
velocity of the star in the direction of the disk rotation (left panel), and of R$_{\rm min}$, the perigalactic distance of the
stellar orbit (right panel). Thick disk stars are shown as filled circles, the low-$\alpha$ halo stars as open symbols, and 
the remaining halo stars as starred symbols. The low-$\alpha$ stars tend to have V close to zero and R$_{\rm min}$ $\leq$ 1 
 kpc. This figure is adapted from \cite[Smiljanic et al. (2009)]{Sm09}.}
   \label{fig4}
\end{center}
\end{figure}

\begin{figure}[t]
\begin{center}
 \includegraphics[width=2.6in]{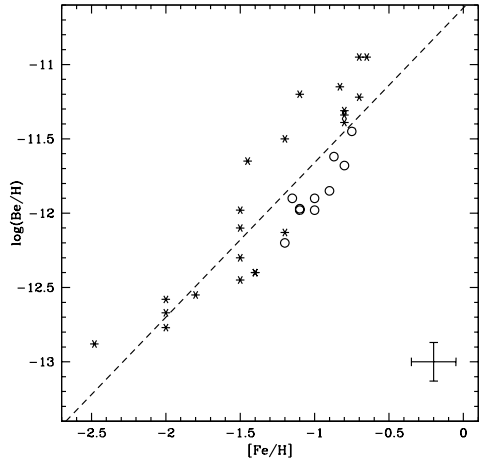} 
 \caption{The beryllium abundances as a function of [Fe/H] only for the halo stars. The low-$\alpha$ stars 
 seem to form a tight and well defined sub-sequence. This figure is adapted from \cite[Smiljanic et al. (2009)]{Sm09}.}
   \label{fig5}
\end{center}
\end{figure}

One remaining question is how does this possible division of the halo stars compare to our current understanding of the 
formation of the Galactic halo? Given that we are analyzing a sample of local halo stars this division is likely not a consequence 
of the inner vs. outer halo dichotomy, as discussed for example by \cite[Carollo et al. (2007, 2009)]{Car07, Car09}. According to 
\cite[Carollo et al. (2007)]{Car07}, the outer halo 
has a metallicity distribution that peaks at [Fe/H] $\sim$ $-$2.20 and dominates the stellar population in distances beyond 15--20 Kpc 
from the Galactic center. The inner halo, on the other hand, has a metallicity distribution that peaks at [Fe/H] = $\sim$ $-$1.60 and dominates 
the stellar population in distances up to 10--15 Kpc. Although we did not check in detail to which of these halo components our sample 
stars belong, we note that they have metallicities between [Fe/H] = $-$2.00 and $-$0.50. The low-$\alpha$ stars in particular have 
$-$1.20 $\leq$ [Fe/H] $\leq$ $-$0.70. We thus believe most of our sample stars are definitely inner halo stars. This implies that our results 
suggest a dichotomy of the inner halo.

There are other literature results that seem to indicate a possible division of the inner halo, from both the observational and the 
theoretical point of view. On the observational side, \cite[Morrison et al. (2009)]{Morr09}, based on the kinematics of a sample of metal-poor stars, conclude 
that the local inner halo seems to divide into two components. One is moderately flattened, has no rotation, has a clumpy distribution in 
energy and angular momentum, and [Fe/H] $<$ $-$1.50. The second is highly flattened, has a small prograde rotation, and $-$1.50 $<$ 
[Fe/H] $<$ $-$1.00. This last component is distinct from the metal-weak thick disk. 

On the theoretical side, \cite[Zolotov et al. (2009)]{Zolotov09} present new simulations of the formation of disk galaxies in a 
$\Lambda$CDM universe. They allow for star formation both in the primary potential well of the galaxy being modelled and in 
dark matter subhalos that are later accreted. They show that the final stellar population in the inner halo of a galaxy formed like 
this has a dual nature. It is composed both by `in situ stars', formed in the inner Galaxy and later displaced to the halo, and by 
accreted stars, formed in the subhalos.

\begin{acknowledgments}
This work has been supported by studentships and fellowships to R.S. by FAPESP (04/13667-4 and 08/55923-8) and CAPES (1521/06-3), and 
by financial support from the ESO DGDF. R.S. also acknowledges financial support from the organizers for his participation in this meeting. 
\end{acknowledgments}

\end{document}